\documentclass{article}
\usepackage{graphicx}
\usepackage{amsmath}
\usepackage{doublespace}
\usepackage{a4}

\input{tcilatex}

\begin{document}

\bigskip

\bigskip

\bigskip

\bigskip

\bigskip

\bigskip

\bigskip

\ \ \ \ \ \ \ \ \ \ \ \ \ \ \ \ \ \ \ \ \ \ \ \ \ \ \ \textbf{Symmetries in
particle and string theories}

\ \ \ \ \ \ \ \ \ \ \ \ \ \ \ \ \ \ \ \ \ \ \ \ \ \ \ \ \ \ \ \ \ \ \ \ \ \
\ \ \ \ \ \ \ \ \ \ \ \ \ \ W. F. Chagas-Filho$^{\dagger }$

\ \ \ \ \ \ \ \ \ \ \ \ \ \ \ Departamento de Fisica, Universidade Federal
de Sergipe, SE, Brazil

\bigskip\ \ \ \ \ \ \ \ \ \ \ \ \ \ \ \ \ \ \ \ \ \ \ \ \ \ \ \ \ \ \ \ \ \
\ \ \ \ \ \ \ \ \ \ \ \ \ \ \ \ \ \ \ \ 

Abstract: We study the space-time invariances of the bosonic relativistic
particle and bosonic relativistic string using general formulations obtained
by incorporating the Hamiltonian constraints into the formalism. We point
out that massless particles and tensionless strings have a larger set of
space-time invariances than their massive and tensionful partners,
respectively. We also show that it is possible to use the reparametrization
invariance of the string formulation we present to reach the classical
conformal equations of motion without the use of two-dimensional Weyl
scalings of the string world sheet. Finally, we show that it is possible to
fix a gauge with an enlarged number of space-time invariances in which every
point of the free tensionless string moves as if it were an almost-free
massless particle. The existence of such a string motion agrees with what is
expected from gauge theory-string duality.

\bigskip\ \ \ \ \ \ \ \ \ \ \ \ \ \ \ \ \ \ \ \ \ \ \ \ \ \ \ \ \ \ \ \ \ \
\ \ \ \ \ \ \ \ \ \ \ \ \ \ 

\bigskip

\bigskip

\bigskip

\bigskip

\bigskip

\bigskip

\bigskip $\dagger $E-mail: wfilho@fisica.ufs.br


\bigskip

\bigskip

\bigskip

\bigskip

\bigskip

\section{Introduction}

\ \ In all fundamental theories of matter, it is necessary to understand
very well the free theory before trying to describe interactions. In this
context, it is sometimes desirable to construct a way along which the
interacting theory could be naturally linked to the free theory, thus
avoiding interactions of being artificially introduced.

One way to investigate the relations between the free and the interacting
theories is to examine the invariances of the corresponding actions. In
interesting cases the free action has a larger number of invariances than
the interacting one and in these cases the appearance of an interaction may
be associated to the breaking of a specific invariance. There is also the
possibility that the breaking of a symmetry causes the appearance of mass
[1].

In this work we are particularly interested in space-time invariances, and
how these invariances manifest themselves in the actions that describe
relativistic objects embedded in the space-time. We focus our attention on
relativistic particles and strings and verify the invariance of the actions
that describe these objects under a general space-time transformation such
as [2] 
\begin{equation}
\delta x^{\mu }=a^{\mu }+\omega ^{\mu \nu }x_{\nu }+\alpha x^{\mu }+(2x^{\mu
}x^{\nu }-\eta ^{\mu \nu }x^{2})b_{\nu }  \tag{1}
\end{equation}
The first two terms on the right in (1) represent a Poincar\'{e}
transformation. The third is a scale transformation and the fourth is a
conformal transformation. $\alpha $ is a constant and $b_{\mu }$ is a
constant vector. The algebra of these space-time transformations is a closed
algebra [2].

At present, the two fundamental theories of matter are Quantum Field Theory
and General Relativity. By the same time, the most promising hope for a
truly unified and finite description of these two fundamental theories is
Superstring Theory [3,4]. This is enough reason to study the dynamics of
relativistic strings. But there is also the important fact that String
Theory was originally discovered as a quantum theory, some time before a
classical string action was first written [4]. String theory has been
developing backwards since then [4], with its geometric structure behind the
classical action still a subject of interest.

In this work we consider only bosonic relativistic strings. We use the
reparametrization invariance of the theory to obtain information about the
classical dynamics. Under reparametrizations, both the bosonic and fermionic
degrees of freedom of the superstring behave as scalars. So, from the
reparametrization invariance point of view, nothing substantially new is
introduced by the fermionic degrees of freedom and we can omit them for
simplicity.

In our approach to the subject we treat the particle and the string as
constrained systems [5] and this allows us to construct classically
equivalent, but more general, particle and string actions that, in the
particular case of strings, have some advantages when compared with the
usual formulations.

In string theory, the most traditional approach to quantization starts by
using the reparametrization invariance of the theory, together with the
two-dimensional Weyl scaling invariance, to impose the conformal gauge. In
this gauge the world sheet metric $h_{\alpha \beta }$ is set equal to the
flat two-dimensional Minkowski metric $\eta _{\alpha \beta }$. The Weyl
scaling symmetry is responsible for the tracelessness of the two-dimensional
energy-momentum tensor $T_{\alpha \beta }$. But the quantum theory gives
rise to an anomaly in the trace of $T_{\alpha \beta }$ and only under very
special circumstances does this anomaly cancel. This is related to the fact
that quantum string theory seems to work only in certain specific space-time
dimensions [3]. In the string formulation we present here, the conformal
equation of motion can be reached using reparametrization invariance only.
No use is made of two-dimensional Weyl invariance.

Another advantage of the string formulation we present is that it can be
used to study the tensionless limit of string theory. The massless limit is
the high-energy limit of particle theory and, as became clear in ref. [6],
the tensionless limit is the high-energy limit of string theory. As was
pointed out in [6], the existence of infinite linear relations between
different string quantum scattering amplitudes with the same momentum, in
the high-energy limit, means that the tensionless string must be more
symmetric than usual strings. But in ref. [6] no mention exists of what this
larger symmetry of tensionless strings could be, or what its physical
origin. This situation was further developed in ref. [7] by the use of a
proposed [8] Ward identity between quantum Green functions. It was shown
that the unknown larger symmetry of the tensionless string is related to the
decoupling of zero-norm states from the theory [7].

Quantum Ward identities in a field theory have their origin in the local
invariances of the corresponding classical action [4]. The evolution of a
string defines a two-dimensional field theory, with the string coordinates $%
x^{\mu }$ as functions of the two parameters used to describe the evolution
surface. On this surface, each of the four space-time transformations in (1)
is a local transformation. In this work we point out that tensionless
strings are invariant under the four transformations in (1), but that
tensionful strings are invariant only under two of them. This is somehow
related to the results in refs. [6] and [7]. We also find that free
tensionless strings are invariant under another local transformation that is
different from the transformations in (1). As we shall see here, these extra
invariances of tensionless strings are natural extensions of extra
invariances of massless particles.

Tensionless strings were originally introduced by Schild in a postumous
paper [9] as strings with a singular world-surface metric tensor, $\det
g^{\alpha \beta }=0$. They are much simpler relativistic objects than usual
strings and attracted attention some time ago [10,11,12, 13] because of the
question if a critical dimension would emerge from its dynamics. No really
definitive answer to this question was arrived at. At the same time,
theoreticians were developing the idea [4] that the larger the gauge group
of a field theory, the larger the possibility that the theory is finite
order by order in a quantum perturbation expansion, and so does not need to
be renormalized. This brought a new interest in the two-dimensional quantum
field theory defined by tensionless strings, which are now being
investigated using different formulations because of its richer symmetries
(see, for instance, refs. [14,15] ). In this work we show that it is
possible to fix a gauge in which each point of the free tensionless string
moves as if it were an almost free massless particle. This was first pointed
out for tensionless superstrings in [11], with each string point moving as
an almost free massless superparticle. This particular possible motion of
tensionless strings, but in the framework of Feynman path integrals, was
also mentioned in [15]. This kind of string motion is expected from gauge
theory-string duality [24,25]. It is also shown here that this particular
gauge-fixed tensionless string, when in free motion, has one extra
invariance when compared with the non gauge-fixed tensionless one, in
complete correspondence with the free massless particle case.

The paper is organized as follows: In section two we consider bosonic
relativistic particle theory. Starting from the usual square-root action we
compute a particle action that allows a transition to the massless limit of
the theory. We investigate the classical equations of motion and the
space-time invariances for both the $m\neq 0$ and $m=0$ cases. We find that
the massless action has three additional invariances which are not present
in the massive action. In section three we extend our procedure to the case
of bosonic relativistic string theory. After a brief introduction to the
subject we show how we can compute a string action that is compatible with
the tensionless limit and which allows us to reach the usual classical
conformal-gauge equations of motion without requiring two-dimensional Weyl
scale invariance. We show that in this formulation the $T=0$ string action
has only two additional space-time invariances when compared with the $T\neq
0$ action. In section four we study in detail the reparametrization
invariance of the string formulation we present and show that it is possible
to fix a gauge in which the $T=0$ string has three extra symmetries when
compared to the $T\neq 0$ string. This creates a link between free massless
relativistic particle theory and this particular sector of free tensionless
relativistic string theory. We present our conclusions in section five.

\section{Relativistic particles}

A relativistic particle describes in space-time a one-parameter trajectory $%
x^{\mu }(\tau )$. The dynamics of the particle must be independent of the
parameter choice. A possible form of the action is the one proportional to
the arc length traveled by the particle and given by 
\begin{equation}
S=-m\int ds=-m\int d\tau \sqrt{-\dot{x}^{2}(\tau )}  \tag{2}  \label{1}
\end{equation}
where $\tau $ is an arbitrary parameter, $m$ is the particle's mass, $%
ds^{2}=-\eta _{\mu \nu }dx^{\mu }dx^{\nu }$, and $\dot{x}^{\mu }=\frac{%
dx^{\mu }}{d\tau }$. Our metric convention is $\eta _{\mu \nu
}=(-1,+1,+1,+1) $ and we use units in which $\hbar =c=1.$

Action (2) is invariant under Poincar\'{e} transformations 
\begin{equation}
\delta x^{\mu }=a^{\mu }+\omega ^{\mu \nu }x_{\nu }  \tag{3}
\end{equation}
with $\omega ^{\mu \nu }=-\omega ^{\nu \mu }$ , and under reparametrizations
of the world-line 
\begin{equation}
\tau \rightarrow \tau ^{\prime }=f(\tau )  \tag{4}
\end{equation}
where $f$ \ is an arbitrary function of $\tau $. The classical equation of
motion for $x^{\mu }(\tau )$ that follows from action (2) implies that 
\begin{equation}
\ddot{x}^{\mu }=0  \tag*{5}
\end{equation}
Action (2) is quite awkward to deal with because of the square root, and it
is also quite restricted because it does not allow a transition to the $m=0$
limit of the theory. The formal problem of defining an action for a massless
particle is naturally solved by treating the particle as a gauge system in
which the gauge invariance is the reparametrization invariance (4).
Nowadays, the most general treatment of a gauge system is within the
constrained Hamiltonian formalism [16].

In the transition to the Hamiltonian formalism action (2) gives the
canonical momentum 
\begin{equation}
p_{\mu }=\frac{m}{\sqrt{-\dot{x}^{2}}}\dot{x}_{\mu }  \tag{6}
\end{equation}
and this momentum gives rise to the primary [5] constraint 
\begin{equation}
\phi =\frac{1}{2}(p^{2}+m^{2})\approx 0  \tag{7}
\end{equation}
We follow Dirac's convention that a constraint is set equal to zero only
after all calculations have been performed. In this sense equation (7) means
that $\phi $ ''weakly'' vanishes. The canonical Hamiltonian that follows
from action (2), $H=p.\dot{x}-L$, identically vanishes. This is a
characteristic feature of reparametrization invariant systems. Dirac's [5]
extended Hamiltonian for the particle is then 
\begin{equation}
H_{E}=H+\lambda \phi =\frac{1}{2}\lambda (p^{2}+m^{2})\approx 0  \tag{8}
\end{equation}
where $\lambda (\tau )$ is a Lagrange multiplier enforcing the constraint $%
\phi $.

The canonical equations of motion obtained from (8) are 
\begin{equation}
\dot{x}^{\mu }=\{x^{\mu },H_{E}\}=\lambda p^{\mu }  \tag{9}
\end{equation}
\begin{equation}
\dot{p}^{\mu }=\{p^{\mu },H_{E}\}=0  \tag{10}
\end{equation}
in which \{ , \} denotes a Poisson bracket. Combining equations (9) and (10)
we obtain 
\begin{eqnarray}
\ddot{x}^{\mu } &=&\dot{\lambda}p^{\mu }+\lambda \dot{p}^{\mu }  \notag \\
&=&\dot{\lambda}p^{\mu }  \TCItag{11}
\end{eqnarray}
In order to have $\ddot{x}^{\mu }=0,$ we must impose the condition $\dot{%
\lambda}=0$. The free particle theory obtained from Hamiltonian (8) is the
one in which $\lambda $ is a constant. But in the general theory, where $%
\lambda $ has a $\tau $-dependence, an interaction proportional to $\dot{%
\lambda}$ is present.

The Lagrangian function that corresponds to the extended Hamiltonian (8) is 
\begin{eqnarray}
L &=&p.\dot{x}-H_{E}  \notag \\
&=&p.\dot{x}-\frac{1}{2}\lambda (p^{2}+m^{2})  \TCItag{12}
\end{eqnarray}
$L$ in equation (12) is defined on an interface between configuration space
and phase space. Using equation (9) to eliminate the $p_{\mu }$ we arrive at
the particle action 
\begin{equation}
S=\frac{1}{2}\int d\tau (\lambda ^{-1}\dot{x}^{2}-\lambda m^{2})  \tag{13}
\end{equation}
Action (13) formally coincides with the ``einbein'' action [3] 
\begin{equation*}
S=\frac{1}{2}\int d\tau (e^{-1}\dot{x}^{2}-em^{2})
\end{equation*}
in which the variable $e(\tau )$ describes the one-dimensional geometry of
the world-line. In order to clarify the role played by $\lambda (\tau )$ in
the theory defined by (13), let us briefly investigate the Hamiltonian
formalism for this action. Action (13) gives the momentum 
\begin{equation}
p_{\mu }=\frac{\dot{x}_{\mu }}{\lambda }  \tag{14}
\end{equation}
and the primary constraint 
\begin{equation}
p_{\lambda }=\frac{\partial L}{\partial \dot{\lambda}}\approx 0  \tag{15}
\end{equation}
The corresponding canonical Hamiltonian is 
\begin{equation}
H=\frac{1}{2}\lambda (p^{2}+m^{2})  \tag{16}
\end{equation}
Incorporating constraint (15) into the formalism we get the extended
Hamiltonian 
\begin{equation}
H_{E}=\frac{1}{2}\lambda (p^{2}+m^{2})+\varsigma p_{\lambda }  \tag{17}
\end{equation}
where $\varsigma $ is a multiplier. Following Dirac's algorithm for
constrained systems we must now require that all constraints be stable. The
stability condition for constraint (15), $\dot{p}_{\lambda }=\{p_{\lambda
},H_{E}\}=0$ is satisfied only if $\phi =\frac{1}{2}(p^{2}+m^{2})\approx 0$.
Requiring now the stability of $\phi ,$ $\ \dot{\phi}=\{\phi ,H_{E}\}=0$, we
find that it is automatically satisfied. The algorithm is completed. No
condition is placed on $\lambda (\tau )$. It remains as an arbitrary
function in action (13). This is a consequence of the fact that $\phi $,
having vanishing Poisson bracket with itself, is a first-class [5]
constraint. Since the $\tau $-derivative of $\lambda $ does not appear in
(13) its equation of motion is a constraint and $\lambda $ can be
eliminated, giving back action (2). Actions (2) and (13) are therefore
classically equivalent.

Action (13) is invariant under the Poincar\'{e} transformation (3) with $%
\delta \lambda =0$ and under the infinitesimal reparametrizations 
\begin{equation}
\delta x^{\mu }=\epsilon \dot{x}^{\mu }  \tag{18a}
\end{equation}
\begin{equation}
\delta \lambda =\frac{d}{d\tau }(\epsilon \lambda )  \tag{18b}
\end{equation}
where $\epsilon (\tau )$ is an arbitrary infinitesimal parameter. The
equation of motion for $x^{\mu }$ that follows from action (13) states that 
\begin{equation}
\frac{d}{d\tau }(\frac{1}{\lambda }\dot{x}^{\mu })=0  \tag{19}
\end{equation}
and again, to select the free theory, we must impose the condition that $%
\dot{\lambda}=0$. The condition that $\lambda $ must be a constant in the
free particle theory has far-reaching consequences. For instance, fixing $%
\lambda =1$ is a fundamental step in the functional quantization procedure
based on action (13) because it precedes [4] the insertion of the
Faddeev-Popov [17] term into the functional integral to get the correct
measure. After exponentiation of the Faddeev-Popov term by the use of
Grassmann variables, the resulting gauge-fixed effective action turns out to
have an important residual global fermionic symmetry called the BRST [18,19]
symmetry.

Action (13), although classically equivalent to, is more general than action
(2) because it allows a transition to the $m=0$ limit of the theory. A
massless relativistic particle can be described by the action 
\begin{equation}
S=\frac{1}{2}\int d\tau \lambda ^{-1}\dot{x}^{2}  \tag{20}
\end{equation}
which is the $m=0$ limit of (13). Action (20) is again invariant under the
Poincar\'{e} transformation (3) with $\delta \lambda =0$ and under the
reparametrizations (18). The equation of motion for $x^{\mu }$ that follows
from action (20) is identical to the equation of motion (19) for the massive
particle. The $m\neq 0$ and $m=0$ particles are governed by the same
classical dynamics. In order to select the free massless theory we must
again impose that $\dot{\lambda}=0$. However, action (20) exhibts
invariances which are not present in action (13).

The massless particle action (20) is invariant under the scale
transformation 
\begin{equation}
\delta x^{\mu }=\alpha x^{\mu }  \tag{21a}
\end{equation}
\begin{equation}
\delta \lambda =2\alpha \lambda  \tag{21b}
\end{equation}
and also invariant under the conformal transformation 
\begin{equation}
\delta x^{\mu }=(2x^{\mu }x^{\nu }-\eta ^{\mu \nu }x^{2})b_{\nu }  \tag{22a}
\end{equation}
\begin{equation}
\delta \lambda =4\lambda x.b  \tag{22b}
\end{equation}
These two invariances are present in the general theory, the one in which
the equation of motion is given by (19) and where there is an arbitrary $%
\tau $-dependence for $\lambda $. They are also present if $\lambda $ has no 
$\tau $-dependence but can suffer arbitrary variations. But there is a
further invariance of action (20) which is present only in the free theory,
where $\dot{\lambda}=0$. Using the equation for free motion, $\ddot{x}^{\mu
}=0$ , it can be verified that the massless action (20) is invariant under
the ``gauge'' transformation 
\begin{eqnarray}
x^{\mu }(\tau ) &\rightarrow &\exp \{\frac{1}{3}\beta (\dot{x}^{2})\}x^{\mu
}(\tau )  \TCItag{23a} \\
\lambda &\rightarrow &\exp \{\frac{2}{3}\beta (\dot{x}^{2})\}\lambda 
\TCItag{23b}
\end{eqnarray}
We call transformation (23) a gauge transformation because $\beta $ is an
arbitrary function of $\dot{x}^{2}$. In the free theory any function $\beta (%
\dot{x}^{2})$ satisfies $\dot{\beta}=0$ and this, together with (23b),
renders action (20) invariant.

The conclusions of this section on relativistic particles are clear. As a
consequence of the first-class property of $\phi $ , the variable $\lambda
(\tau )$ remains as an arbitrary function in action (13). $\lambda $ acts as
a gauge function that can be used to select the free sector of relativistic
particle theory. As we also saw, the massless limit of particle theory is
more symmetric than the massive one, with the free $m=0$ sector having three
extra invariances if compared to the $m\neq 0$ sector.

It is well known [2] that the presence of scale invariance automatically
implies the presence of conformal invariance. We are then led to the idea
that the breaking of the space-time scale invariance (21), and consequently
the breaking of the space-time conformal invariance (22), are related to the
appearance of a non-vanishing particle mass in action (13). This is true in
the general and in the free theory. Following this same reasoning, the
breaking of the ``gauge'' invariance (23) must be related to the appearance
of an interaction in the massless theory. This interaction is presumably of
gravitational origin, and is related to a non-vanishing $\tau $-derivative
of $\lambda $. The free massive theory does not have the ``gauge''
invariance (23) because the mass term spoils the invariance of action (13)
under transformation (23). In the next two sections we show how this
physical picture can be extended to string theory.

\section{Relativistic strings}

Strings are higher-dimensional extensions of the particle concept. As a
consequence of its evolution, the string traces out a world sheet in
space-time. In the form originally advocated by Nambu [20] and Goto [21],
the action for a string is simply proportional to the area of its world
sheet. Mathematically, one formula for the area of a sheet embedded in
space-time is 
\begin{equation}
S=T\int d\tau d\sigma \sqrt{\dot{x}^{2}\acute{x}^{2}-(\dot{x}.\acute{x})^{2}}
\tag{24}
\end{equation}
in which $x^{\mu }=x^{\mu }(\tau ,\sigma )$ are the coordinates of the
string, given as $D$ functions of the parameters that describe the world
surface, and primes denote derivatives with respect to $\sigma $. $T$ is a
constant of proportionality, required to make the action dimensionless. $T$
must have dimension of $(length)^{-2}$, or $(mass)^{2}$, to leave a
dimensionless action. It can be shown [3, 22] that $T$ is actually the
tension in the string. The actual value of $T$ may be treated as a free
parameter in the theory and corresponds to an energy scale. With the action
of the string taken to be the area of its world sheet, the solutions of the
classical equations of the free string are the world sheets of minimal (or
at least extremal) area. This generalizes the fact that the trajectories of
the free particle are geodesics, or curves of minimal length.

It is difficult to work with action (24) because it is highly nonlinear and
especially because of the square root. An equivalent, but more convenient
form of the action can be written if we introduce in addition to $x^{\mu
}(\tau ,\sigma )$ a new variable $h_{\alpha \beta }$, which will be a metric
tensor of the string world sheet. This more convenient action is [23] 
\begin{equation}
S=-\frac{T}{2}\int d\tau d\sigma \sqrt{h}h^{\alpha \beta }\eta _{\mu \nu
}\partial _{\alpha }x^{\mu }\partial _{\beta }x^{\nu }  \tag{25}
\end{equation}
Here $\sqrt{h\text{ }}$ is the square root of the absolute value of the
determinant of $h_{\alpha \beta }$ and $h^{\alpha \beta }$ is the inverse of 
$h_{\alpha \beta }$. Action (25) is the standard form for coupling $D$
massless scalar fields $x^{\mu }$ to (1+1)-dimensional gravity [3]. Since
the derivatives of $h_{\alpha \beta }$ do not appear in (25) its equation of
motion is a constraint and $h_{\alpha \beta }$ can be integrated out giving
back action (24).

Either (24) or (25) is invariant under general coordinate transformations of
the string world sheet $\tau \rightarrow \tau ^{\prime }(\tau ,\sigma ),$ $%
\sigma \rightarrow \sigma ^{\prime }(\tau ,\sigma )$. This reparametrization
invariance is essential for solving the minimal surface equations derived
from (25). The $2\times 2$ symmetric tensor $h_{\alpha \beta }$ has three
independent components. By a suitable choice of new parameters $\tau
^{\prime }$ and $\sigma ^{\prime }$ any two of the three independent
components of $h_{\alpha \beta }$ can be eliminated. This leaves only one
independent component. However, there is one more local two-dimensional
symmetry of the string action (25). There is a local Weyl scaling of the
metric 
\begin{equation}
h_{\alpha \beta }\rightarrow \Lambda (\tau ,\sigma )h_{\alpha \beta } 
\tag{26}
\end{equation}
which leaves the factor $\sqrt{h}h^{\alpha \beta }$ invariant. The
reparametrization invariance together with the Weyl scaling can then be used
to gauge away all the three independent $\ h_{\alpha \beta }$ components by
imposing the conformal gauge $h_{\alpha \beta }=\eta _{\alpha \beta }$ ,
where $\eta _{\alpha \beta }$ is the two-dimensional flat space metric. In
this gauge action (25) becomes 
\begin{equation}
S=-\frac{T}{2}\int d\tau d\sigma \eta ^{\alpha \beta }\eta _{\mu \nu
}\partial _{\alpha }x^{\mu }\partial _{\beta }x^{\nu }  \tag{27}
\end{equation}
The equation of motion derived from action (27) is the free two-dimensional
linear wave equation 
\begin{equation}
\frac{\partial ^{2}x^{\mu }}{\partial \tau ^{2}}-\frac{\partial ^{2}x^{\mu }%
}{\partial \sigma ^{2}}=0  \tag{28}
\end{equation}
The solutions of the wave equation (28) are well-known [3]. These solutions
form the basis of the ``Old Covariant First Quantized'' formalism for
bosonic relativistic strings [3].

Now, while the string action (24) is obviously the higher-dimensional
extension of the particle action (2), the string extension of the particle
action (13) is still lacking. This is because action (25) can not be used to
study the tensionless limit of string theory, as action (13) was used to
study the massless limit of particle theory. We see that it is necessary to
compute a string action that allows a transition to the $T=0$ limit. To
parallel our treatment of the relativistic particle as close as possible we
return to the metric convention $\eta _{\mu \nu }=-1,+1,...,+1$ and write
the Nambu-Goto action as 
\begin{equation}
S=-T\int d\tau d\sigma \sqrt{-g}  \tag{29a}
\end{equation}
where 
\begin{eqnarray}
g &=&\det g_{\alpha \beta }  \TCItag{29b} \\
g_{\alpha \beta } &=&\eta _{\mu \nu }\partial _{\alpha }x^{\mu }\partial
_{\beta }x^{\nu }  \TCItag{29c}
\end{eqnarray}
$g_{\alpha \beta }$ is the two-dimensional metric induced on the string
world sheet as a result of its embedding in space-time. In the transition to
the Hamiltonian formalism the Nambu-Goto action (29) gives the canonical
momentum 
\begin{equation}
p_{\mu }=-T\sqrt{-g}g^{0\alpha }\partial _{\alpha }x_{\mu }  \tag{30}
\end{equation}
and this momentum gives rise to the primary constraints 
\begin{eqnarray}
\Phi &=&\frac{1}{2}(p^{2}+T^{2}\acute{x}^{2})\approx 0  \TCItag{31a} \\
\Phi _{1} &=&p.\acute{x}\approx 0  \TCItag{31b}
\end{eqnarray}
Due to the reparametrization invariance of the theory the canonical
Hamiltonian density vanishes and Dirac's Hamiltonian density for the
Nambu-Goto string is 
\begin{eqnarray}
H_{E} &=&H+\lambda \Phi +\lambda _{1}\Phi _{1}  \notag \\
&=&\lambda \Phi +\lambda _{1}\Phi _{1}  \TCItag{32}
\end{eqnarray}
where $\lambda (\tau ,\sigma )$ and $\lambda _{1}(\tau ,\sigma )$ are
Lagrange multipliers for the constraints $\Phi $ and $\Phi _{1}$. The
Lagrangian density corresponding to (32) is 
\begin{eqnarray}
L &=&p.\dot{x}-\lambda \Phi -\lambda _{1}\Phi _{1}  \notag \\
&=&p.\dot{x}-\frac{\lambda }{2}(p^{2}+T^{2}\acute{x}^{2})-\lambda _{1}p.%
\acute{x}  \TCItag{33}
\end{eqnarray}
The string Lagrangian (33) is defined in an extended configuration space.
But we can return to the usual configuration space by integrating the
momenta. The solution of the equation of motion for $p_{\mu }$\ is 
\begin{equation}
p_{\mu }=\frac{1}{\lambda }(\dot{x}_{\mu }-\lambda _{1}\acute{x}_{\mu }) 
\tag{34}
\end{equation}
Inserting this result back into (33) we obtain the string action 
\begin{equation}
S=\frac{1}{2}\int d\tau d\sigma \lbrack \lambda ^{-1}(\dot{x}-\lambda _{1}%
\acute{x})^{2}-\lambda T^{2}\acute{x}^{2}]  \tag{35}
\end{equation}
Action (35) is the higher-dimensional extension of the particle action (13).
It is Poincar\'{e} invariant with $\delta \lambda =0,$ $\delta \lambda
_{1}=0 $. It is also reparametrization invariant, as will be explicitly
checked in the next section. If we eliminate $\lambda $ and $\lambda _{1}$
through their equations of motion we recover the Nambu-Goto action (24).
Action (35) is then classically equivalent to the Nambu-Goto action.
However, it is more general than actions (24) and (25) because it allows a
transition to the $T=0 $ limit.

If we construct the Hamiltonian formulation for action (35) we will find
that the two-dimensional fields $\lambda $ and $\lambda _{1}$ remain as
arbitrary functions in the theory. This is because constraints (31) are
first-class constraints obeying the Poisson bracket gauge algebra 
\begin{eqnarray}
\{\phi (\tau ,\sigma ),\phi (\tau ,\sigma ^{\prime })\} &=&T^{2}[\phi
_{1}(\tau ,\sigma )+\phi _{1}(\tau ,\sigma ^{\prime })]\delta ^{\prime
}(\sigma -\sigma ^{\prime })  \TCItag{36a} \\
\{\phi (\tau ,\sigma ),\phi _{1}(\tau ,\sigma ^{\prime })\} &=&[\phi (\tau
,\sigma )+\phi (\tau ,\sigma ^{\prime })]\delta ^{\prime }(\sigma -\sigma
^{\prime })  \TCItag{36b} \\
\{\phi _{1}(\tau ,\sigma ),\phi _{1}(\tau ,\sigma ^{\prime })\} &=&[\phi
_{1}(\tau ,\sigma )+\phi _{1}(\tau ,\sigma ^{\prime })]\delta ^{\prime
}(\sigma -\sigma ^{\prime })  \TCItag{36c}
\end{eqnarray}
We may, for instance, use the arbitrariness of $\lambda $ and $\lambda _{1}$
to make the identification 
\begin{eqnarray}
-T\sqrt{h}h^{00} &=&\frac{1}{\lambda }  \TCItag{37a} \\
-T\sqrt{h}h^{01} &=&-\frac{\lambda _{1}}{\lambda }  \TCItag{37b} \\
-T\sqrt{h}h^{11} &=&\frac{\lambda _{1}^{2}}{\lambda }-\lambda T^{2} 
\TCItag{37c}
\end{eqnarray}
Action (35) can then be rewritten as

\begin{equation*}
S=-\frac{T}{2}\int d\tau d\sigma \sqrt{h}h^{\alpha \beta }\partial _{\alpha
}x.\partial _{\beta }x
\end{equation*}
which is identical to (25). In equations (37), three independent metric
factors are expressed in terms of two independent arbitrary functions $%
\lambda $ and $\lambda _{1}$. This means that in the string formulation (35)
the conformal-gauge equation of motion (28) can be reached by use of
reparametrization invariance only. This can be explicitly verified. The
classical equation of motion for $x_{\mu }$ that follows from action (35) is 
\begin{equation}
\frac{\partial }{\partial \tau }(\frac{1}{\lambda }\dot{x}_{\mu }-\frac{%
\lambda _{1}}{\lambda }\acute{x}_{\mu })+\frac{\partial }{\partial \sigma }[-%
\frac{\lambda _{1}}{\lambda }\dot{x}_{\mu }+(\frac{\lambda _{1}^{2}}{\lambda 
}-\lambda T^{2})\acute{x}_{\mu }]=0  \tag{38}
\end{equation}
Now, choosing $\lambda =1,$ if we want to reproduce the conformal-gauge
equation of motion, we have to satisfy the condition $\lambda
_{1}^{2}-T^{2}=-1,$ for which the solutions are $\lambda _{1}=\pm \sqrt{%
T^{2}-1}$. Using the positive value of $\lambda _{1}$ in the first term of
(38), the negative value of $\lambda _{1}$ in the second term, and the fact
that the partial derivatives with respect to $\tau $ and $\sigma $ commute,
equation of motion (38) reduces to 
\begin{equation}
\frac{\partial ^{2}x_{\mu }}{\partial \tau ^{2}}-\frac{\partial ^{2}x_{\mu }%
}{\partial \sigma ^{2}}=0  \tag{39}
\end{equation}
which is identical to equation (28). In the special case of equation of
motion (38) reparametrization invariance allows us to make three independent
gauge choices. In the ``Old Covariant First Quantized'' [3] formalism based
on the gauge-fixed classical action (27) the trace anomaly is an
inconsistency because it implies the non-validity of the wave equation (28),
upon which all the formalism is based. The anomaly must then be eliminated,
which can only be completely done in $D=26$. Here this problem does not
exist because no use is made of two-dimensional Weyl scale invariance to
reach the wave equation (39). The anomaly must manifest itself in another
way.

Taking the limit $T=0$ in action (35) we obtain the string action 
\begin{equation}
S=\frac{1}{2}\int d\tau d\sigma \lambda ^{-1}(\dot{x}-\lambda _{1}\acute{x}%
)^{2}  \tag{40}
\end{equation}
With the identification 
\begin{eqnarray}
g^{00} &=&\frac{1}{\lambda }  \TCItag{41a} \\
g^{01} &=&-\frac{\lambda _{1}}{\lambda }  \TCItag{41b} \\
g^{11} &=&\frac{\lambda _{1}^{2}}{\lambda }  \TCItag{41c}
\end{eqnarray}
action (40) can be rewritten as 
\begin{equation}
S=\frac{1}{2}\int d\tau d\sigma g^{\alpha \beta }\partial _{\alpha
}x.\partial _{\beta }x  \tag{42}
\end{equation}
which is identical to the one proposed by Schild [9]. Notice that using
equations (41) we have $\det g^{\alpha \beta }=0$.

Action (40) is invariant under Poincar\'{e} transformations and, as will be
checked in the next section, also reparametrization invariant. The equation
of motion for $x_{\mu }$ that follows from action (40) is the $T=0$ limit of
equation (38), namely, 
\begin{equation}
\frac{\partial }{\partial \tau }(\frac{1}{\lambda }\dot{x}_{\mu }-\frac{%
\lambda _{1}}{\lambda }\acute{x}_{\mu })+\frac{\partial }{\partial \sigma }(-%
\frac{\lambda _{1}}{\lambda }\dot{x}_{\mu }+\frac{\lambda _{1}^{2}}{\lambda }%
\acute{x}_{\mu })=0  \tag{43}
\end{equation}
Equation (43) can also be reduced to the conformal wave equation. Choosing
again $\lambda =1$, the possible values of $\lambda _{1}$ are now $\lambda
_{1}=\pm \sqrt{-1}=\pm i$ . Using $\lambda _{1}=i$ in the first term, and $%
\lambda _{1}=-i$ in the second term, equation (43) becomes the conformal
equation of motion (39). We see that in the conformal gauge both the $T\neq
0 $ and $T=0$ strings satisfy free two-dimensional linear wave equations. In
this gauge they are then governed by the same classical dynamics. But, in
general, the $T\neq 0$ \ and \ $T=0$ \ dynamics are different because the
equations of motion (38) and (43) differ by a term proportional to $T$ .

The $T=0$ action (40) has invariances which are not shared by the $T\neq 0$
action (35). Action (40) is invariant under the scale transformation 
\begin{equation}
\delta x^{\mu }=\alpha x^{\mu }  \tag{44a}
\end{equation}
\begin{equation}
\delta \lambda =2\alpha \lambda  \tag{44b}
\end{equation}
\begin{equation}
\delta \lambda _{1}=0  \tag{44c}
\end{equation}
and also invariant under the conformal transformation 
\begin{equation}
\delta x^{\mu }=(2x^{\mu }x^{\nu }-\eta ^{\mu \nu }x^{2})b_{\nu }  \tag{45a}
\end{equation}
\begin{equation}
\delta \lambda =4\lambda x.b  \tag{45b}
\end{equation}
\begin{equation}
\delta \lambda _{1}=0  \tag{45c}
\end{equation}
The breaking of the $D$-dimensional scale invariance and conformal
invariance must then be related to the appearance of a non-vanishing tension
value.

As we saw above, the $T=0$ string action (40) has two additional space-time
invariances when compared with the $T\neq 0$ action (35): scale invariance
and conformal invariance. But, as we saw in the last section, the $m=0$
particle action (20) has three additional space-time invariances when
compared with the $m\neq 0$ particle action (13). Here action (40) is not
invariant under the string extension of the particle transformation (23). In
the next section we show that it is possible to fix a gauge in which the $%
T=0 $ string is invariant under a transformation that is the extension of
the $m=0$ particle transformation (23). This creates a link between these
two particular relativistic objects, in the sense that they have identical
extra space-time invariances when compared with their usual ($m\neq 0$ and $%
T\neq 0 $ ) partners.

\section{Gauge-fixed strings}

The $\lambda $ variations in equations (44b) and (45b) have the same
mathematical structure as the particle variations (21b) and (22b). But
equations (44c) and (45c) indicate that $\lambda _{1}$ is useless in
establishing space-time scale and conformal invariances. To clarify this
situation we study the reparametrization invariance of the $T\neq 0$ string
action (35).

Under the infinitesimal reparametrizations of the world-surface coordinates $%
\xi ^{\alpha }=(\tau ,\sigma )$ given by 
\begin{equation}
\xi ^{\prime \alpha }=\xi ^{\alpha }+\epsilon ^{\alpha }  \tag{46}
\end{equation}
a generic action functional 
\begin{equation}
S[\phi ]=\int d^{2}\xi L(\phi ,\partial _{\alpha }\phi )  \tag{47}
\end{equation}
varies as 
\begin{eqnarray}
\delta S &=&\int d^{2}\xi ^{\prime }L(\phi ^{\prime }(\xi ^{\prime
}),\partial _{\alpha ^{\prime }}\phi ^{\prime }(\xi ^{\prime }))-\int
d^{2}\xi L(\phi (\xi ),\partial _{\alpha }\phi (\xi ))  \notag \\
&=&\int d^{2}\xi \Delta L  \TCItag{48}
\end{eqnarray}
where the variation $\Delta L$ is given by 
\begin{equation}
\Delta L=\partial _{\alpha }\epsilon ^{\alpha }L+\frac{\partial L}{\partial
\phi }\delta \phi +\frac{\partial L}{\partial \phi _{,\alpha }}\delta \phi
_{,\alpha }  \tag{49}
\end{equation}
and 
\begin{equation}
\delta \phi (\xi )=\phi ^{\prime }(\xi ^{\prime })-\phi (\xi )  \tag{50}
\end{equation}
is the total variation of the generic field $\phi (\xi ).$ The string
coordinates $x^{\mu }(\tau ,\sigma )$ are space-time vectors but behave as $%
D $ scalar fields from the world-surface point of view. Under
reparametrizations of the world-surface coordinates, their derivatives vary
as 
\begin{equation}
\delta (\partial _{\alpha }x^{\mu })=-\partial _{\alpha }\epsilon ^{\beta
}\partial _{\beta }x^{\mu }  \tag{51}
\end{equation}
Under these conditions, the $T\neq 0$ string action (35) is
reparametrization invariant for 
\begin{eqnarray}
\delta \lambda &=&(\epsilon _{1}^{\prime }-\dot{\epsilon}_{0})\lambda
-\lambda \lambda _{1}\epsilon _{0}^{\prime }  \TCItag{52a} \\
\delta \lambda _{1} &=&(\epsilon _{1}^{\prime }-\dot{\epsilon}_{0})\lambda
_{1}+\dot{\epsilon}_{1}-\epsilon _{0}^{\prime }\lambda _{1}^{2}-T^{2}\lambda
^{2}\epsilon _{0}^{\prime }  \TCItag{52b}
\end{eqnarray}
Reparametrization invariance of the $T=0$ string action (40) is obtained by
taking this limit in equations (52) above. In particular, equation (52b)
shows that there is enough reparametrization freedom \ to impose the gauge $%
\lambda _{1}(\tau ,\sigma )=0$. In this gauge action (35) simplifies to 
\begin{equation}
S=\frac{1}{2}\int d\tau d\sigma (\lambda ^{-1}\dot{x}^{2}-\lambda T^{2}%
\acute{x}^{2})  \tag{53}
\end{equation}
Constructing the Hamiltonian formalism for action (53) we find the
constraint that the canonical momentum conjugate to $\lambda $ must vanish.
The stability of this constraint requires $\phi $ of equation (31a) as a
secondary constraint, and the stability of $\phi $ requires $\phi _{1}$ of
equation (31b) as a terciary one. The inherent string structure, together
with the gauge algebra (36), are preserved in action (53).

The classical equation of motion that follows from action (53) is 
\begin{equation}
\frac{\partial }{\partial \tau }(\frac{1}{\lambda }\dot{x}^{\mu })-T^{2}%
\frac{\partial }{\partial \sigma }(\lambda \acute{x}^{\mu })=0  \tag{54}
\end{equation}
We can now use the residual reparametrization invariance to choose $\lambda
=1$ and get the wave equation 
\begin{equation}
\frac{\partial ^{2}x^{\mu }}{\partial \tau ^{2}}-T^{2}\frac{\partial
^{2}x^{\mu }}{\partial \sigma ^{2}}=0  \tag{55}
\end{equation}
Now, if we set $T=1,$ we return to the conformal equation of motion (39).
The general equation of motion (54) shows again that when $T=0$ the string
dynamics is modified. In particular, equation (55) shows that when $T=0$ \
the string is no longer described by a wave equation.

Setting $T=0$ in (53), we obtain the string action 
\begin{equation}
S=\frac{1}{2}\int d\tau d\sigma \lambda ^{-1}\dot{x}^{2}  \tag{56}
\end{equation}
To be consistent, action (56) must be complemented with the constraints 
\begin{eqnarray}
\phi &=&p^{2}\approx 0  \TCItag{57a} \\
\phi _{1} &=&p.\acute{x}\approx 0  \TCItag{57b}
\end{eqnarray}
which are the $T=0$ limits of constraints (31). The classical equation of
motion for $x^{\mu }$ that follows from the string action (56) is the $T=0$
limit of equation (54), namely 
\begin{equation}
\frac{\partial }{\partial \tau }(\frac{1}{\lambda }\dot{x}^{\mu })=0 
\tag{58}
\end{equation}
If we now choose $\lambda =cons\tan t$ we find that under this condition 
\begin{equation}
\frac{\partial ^{2}x^{\mu }}{\partial \tau ^{2}}=0  \tag{59}
\end{equation}
and every point $\sigma $ of the tensionless string moves as if it were an
almost free massless particle. These massless particles are not completely
free but are held together in a string structure in consequence of the
presence of constraint (57b). Action (56) has an extra space-time invariance
when compared with the $T=0$ action (40).

Action (56) is invariant under the scale transformation 
\begin{equation}
\delta x^{\mu }=\alpha x^{\mu }  \tag{60a}
\end{equation}
\begin{equation}
\delta \lambda =2\alpha \lambda  \tag{60b}
\end{equation}
and also invariant under the conformal transformation 
\begin{equation}
\delta x^{\mu }=(2x^{\mu }x^{\nu }-\eta ^{\mu \nu }x^{2})b_{\nu }  \tag{61a}
\end{equation}
\begin{equation}
\delta \lambda =4\lambda x.b  \tag{61b}
\end{equation}

When the $T=0$ string satisfies the equation of motion (59), action (56) is
also invariant under the ``gauge'' transformation 
\begin{eqnarray}
x_{\mu }(\tau ,\sigma ) &\rightarrow &\exp \{\frac{1}{3}\gamma (\dot{x}%
^{2})\}x_{\mu }(\tau ,\sigma )  \TCItag{62a} \\
\lambda &\rightarrow &\exp \{\frac{2}{3}\gamma (\dot{x}^{2})\}\lambda 
\TCItag{62b}
\end{eqnarray}
with $\gamma $ an arbitrary function. Transformation (62) is the string
extension of the massless particle transformation (23).

\section{Conclusion}

In this work we treat the relativistic bosonic particle and the relativistic
bosonic string as constrained Hamiltonian systems with first-class
constraints only. After the elimination of the canonical momenta using their
equations of motion we obtain a particle action that allows transition to
the massless limit and a string action that gives the classical conformal
wave equation of motion without use of two-dimensional Weyl scalings an
which allows transition to the tensionless limit. It was shown that the $m=0$
particle action has three extra space-time invariances if compared to the $%
m\neq 0$ action.

It was also shown that there exists a $T=0$ string that can be required to
obey the classical conformal wave equation and that has two extra space-time
invariances if compared to his $T\neq 0$ partner.

Finally, we pointed out that there also exists a simpler, gauge-fixed ,
consistent $T\neq 0$ string which can be required to satisfy the classical
conformal wave equation but that his corresponding $T=0$ string can not.
This $T=0$ \ string obeys particle-like equations of motion which are
consistent only if $p.\acute{x}=0$. This particular string can have three
extra space-time invariances if compared to his $T\neq 0$ version. We
interpret it as an intermediate state between a string-like ordered system
of particles and a $T\neq 0$ string. This particular string seems to agree
with the observation of Sundborg [24] and Witten [25], based on duality
between gauge theory and string theory, that when the gauge theory is in the
weak coupling regime the string tension effectively tends to zero.

\end{document}